\documentclass[reprint,twocolumn,prx,superscriptaddress,longbibliography]{revtex4-1}

\usepackage{graphicx}
\usepackage{color}
\usepackage{amssymb}
\usepackage{amsfonts}
\usepackage{amsmath}
\usepackage{bm}
\usepackage{bbm}
\usepackage{lipsum}

\begin{document}

\title{Dynamical vortex transitions in a gate-tunable Josephson junction array}

\author{C.~G.~L.~B\o{}ttcher}\thanks{Present Address: Department of Applied Physics, Yale University, New Haven, CT, 06520, USA. \\
Email: charlotte.boettcher@yale.edu}
\affiliation{Center for Quantum Devices, Niels Bohr Institute, University of Copenhagen, 2100 Copenhagen, Denmark}
\author{F.~Nichele}\thanks{Present Address: IBM Research Laboratory, Z\"urich, Z\"urich, Switzerland}
\affiliation{Center for Quantum Devices, Niels Bohr Institute, University of Copenhagen, 2100 Copenhagen, Denmark}
\author{J.~Shabani}\thanks{Present Address: New York University, New York, NY 10003, USA}
\affiliation{California NanoSystems Institute, University of California, Santa Barbara, CA 93106, USA}
\author{C.~J.~Palmstr\o{}m}
\affiliation{California NanoSystems Institute, University of California, Santa Barbara, CA 93106, USA}
\affiliation{Department of Electrical Engineering, University of California, Santa Barbara, CA 93106, USA}
\affiliation{Materials Department, University of California, Santa Barbara, CA 93106, USA}
\author{C.~M.~Marcus}
\affiliation{Center for Quantum Devices, Niels Bohr Institute, University of Copenhagen, 2100 Copenhagen, Denmark}

\date{\today}

\begin{abstract}
{\footnotesize We explore vortex dynamics in a two-dimensional Josephson junction array of micron-size superconducting islands fabricated from an epitaxial Al/InAs superconductor-semiconductor heterostructure, with a global top gate controlling Josephson coupling and vortex pinning strength. With  applied dc current, minima of differential resistance undergo a transition, becoming local maxima at integer and half-integer flux quanta per plaquette, $f$. The zero-field transition from the superconducting phase is split, but unsplit for the anomalous metal phase, suggesting that pinned vortices are absent or sparse in the superconducting phase, and abundant but frozen in the anomalous metal. The onset of the transition is symmetric around $f=1/2$ but skewed around $f=1$, consistent with a picture of dilute vortices/antivortices on top of a checkerboard ($f = 1/2$) or uniform array of vortices ($f = 1$). Transitions show good scaling but with exponents that differ from Mott values obtained earlier.  Besides the skewing at $f=1$, transitions show an overall even-odd pattern of skewing around integer $f$ values, which we attribute to vortex commensuration in the square array leading to symmetries around half-integer $f$.
}
\end{abstract}
\maketitle 
\indent 

\section{Introduction}
Two-dimensional (2D) superconductivity in thin films and Josephson junction arrays (JJAs) reveal complex classical and quantum phase transitions and rich dynamics that depend on competing energy scales and coherence, including the extensively studied superconductor-insulator transition (SIT) \cite{Jaeger1989, Fisher1991, Lee1990, Goldman2010, Gantmakher2010, Vladimir2012}, which fits into a framework of quantum critical phenomena exhibiting universal scaling \cite{Mason1999, Yazdani1995, Steiner2005, Bollinger2011, Schneider2012}. New classes of materials have extended SIT studies to include strictly 2D materials \cite{Allain2012,Tamir2019,Fatemi2018}, hybrid super-semi heterostructures \cite{Boettcher2018,Vaitieknas2020,Tosato.2022,Boettcher.2022}, and high-temperature superconductors \cite{Yang2019}, including a transition to a topological insulating phase \cite{Wu2018}. An apparent metallic phase, with saturating low-temperature finite resistance, is observed in many of these systems \cite{Kapitulnik2019}, which is incompatible with simple universal scaling.

JJAs enrich the landscape by allowing controlled Coulomb interaction and frustration due to magnetic flux commensuration \cite{Newrock1999, Fazio2001}. Coulomb charging of Josephson-coupled islands makes the classical XY spin system into a quantum problem, with phase and charge on the islands acting as conjugate variables with an uncertainty relation. Charging also introduces a new type of disorder in the form of a random offset on the island. The periodicity of the array adds complexity in the form of a Hofstadter-like spectrum \cite{Teitel.1983, Lankhorst.2018b}, and possible spin glass phases \cite{Vinokur.1987,Spivak.1991,Phillips.2003}. The discrete structure of a JJA also results in periodic pinning of vortices and antivortices, whose unbinding and free motion at finite temperature is described at zero magnetic field by a  Berzinskii-Kosterlitz-Thouless (BKT) phase transition \cite{Halperin1979, Lobb.1983, Newrock1999, Fazio2001, Gantmakher2010}, as investigated recently in this system \cite{Boettcher.2022}. 

 Collective pinning of vortices resulting in a zero resistance state, which has been mapped onto Mott insulator of frozen vortices \cite{Nelson1993}, was investigated experimentally \cite{Poccia2015, Lankhorst2018, Mironov.2020, Rezvani.2020, Rezvani.2020b, Pei.2022}, showing good agreement with theory, including scaling \cite{Granato.2018,Granato.2019}. In this picture, the zero-resistance state of the JJAs near nonzero integer and half integer flux quanta per plaquette, denoted frustration, $f$, is described as vortices pinned by a combination of the periodic array potential and, importantly, {\it collective} pinning due to other vortices. Individual (noncollective) vortex pinning in the binding potential of the array has been thoroughly modelled in metallic JJAs \cite{Lobb.1983, Rzchowski1990} and for intentional pinning \cite{Berdiyorov.2005} and antipinning sites \cite{Berdiyorov.2008}, patterned to prevent dissipation in superconducting films \cite{Eley.2021}. We note that the Mott insulator in JJAs is closely related to the theoretical Mott insulator state at weak disorder in the 2D Bose Hubbard model. For stronger disorder the Bose Hubbard model shows a glass phase associated with individual vortex pinning \cite{POLLET2013,Yao2014}.

Here, we investigate a dynamic transition from the superconducting to the resistive state driven by an applied dc current, including scaling analysis of differential resistance at the transition. Previous experiments ~\cite{Poccia2015, Lankhorst2018, Mironov.2020, Rezvani.2020, Rezvani.2020b, Pei.2022} and theory \cite{Granato.2018,Granato.2019} interpreted the flux-dependent dynamic transition in terms of a dynamic Mott transition based on a scaling analysis that yielded exponents consistent with a Mott transition. For the gate-tuned semiconductor-based JJA investigated here, both the Josephson coupling, $E_J$ and the vortex pinning potential, $E_B$, are tuned by a gate voltage \cite{Boettcher2018, Boettcher.2022}. 

Examining these transition as a dynamical phase transition, we find reasonable scaling, though with nonuniversal scaling exponents that differ from previous experiments \cite{Poccia2015, Lankhorst2018, Pei.2022}. We also find different exponents on the low-field and high-field sides of the transitions. This is discussed in  (Sec.~\ref{scaling}). We conclude that our transition is not a simple Mott transition in the same universality class as previously reported. This may be due to the geometry of our structure or weaker vortex pinning in the InAs/Al system. We note that earlier studies of similar dynamic transitions were interpreted in terms of depinning of vortices from the array potential rather than as a Mott transition \cite{Benz.1990, Jiang.2004}. 

Beyond scaling, several new features of dynamical vortex transitions are presented, elucidating the underlying vortex structure. First, the dip-to-peak transition from the fully superconducting state is {\it split} at zero magnetic field, $f=0$, unlike the cases of $f=$~1/2, 1, and other integer $f$. We interpret the split peak as reflecting absent or sparse vortices at $f=0$, unlike at nonzero $f$, consistent with a BKT picture at $f=0$ where vortex-antivortex pairs annihilate at low temperature. 

Importantly, we find that when the system is tuned (by gate voltage) into the anomalous metal phase \cite{Boettcher2018}, an {\it unsplit} peak at $f=0$ is instead observed. This suggests that remnant unpaired vortices and antivortices, absent in the low-temperature low-current superconducting state, are abundant at $f=0$ in the anomalous metal under otherwise similar conditions. The size of the splitting at $f=0$ in the superconducting state is found to depend on $I_{\rm dc}$, reflecting the vortex density needed to support a dynamical transition. This is discussed in Sec.~\ref{split}. 

Second, the evolution of the dip-to-peak transition  with increasing $I_{\rm dc}$ is found to be symmetric about $f=1/2$ but highly asymmetric about $f=1$ and larger integers. The right-left symmetry around $f=1/2$ follows from the underlying checkerboard vortex configuration \cite{Franz.1995, Lankhorst.2018b}: excess vortices are attracted to unfilled sites while deficit vortices, equivalent to antivortices, are attracted to filled sites. The situation at $f=1$ is quite different and naturally asymmetric: the array is full, contains one vortex per site. Any excess vortex is repelled at each site while any deficit, or antivortex, is attracted to each site, where it can annihilate. Higher integer $f$ values mirror the asymmetry at $f=1$ about the intervening half-integer symmetry point. This is discussed in Sec.~\ref{evenodd}. 

\section{Hybrid Array and Dynamical Vortex Transition}
\label{device}

\begin{figure*}[t]
    \centering
	\includegraphics[width= 6.4 in]{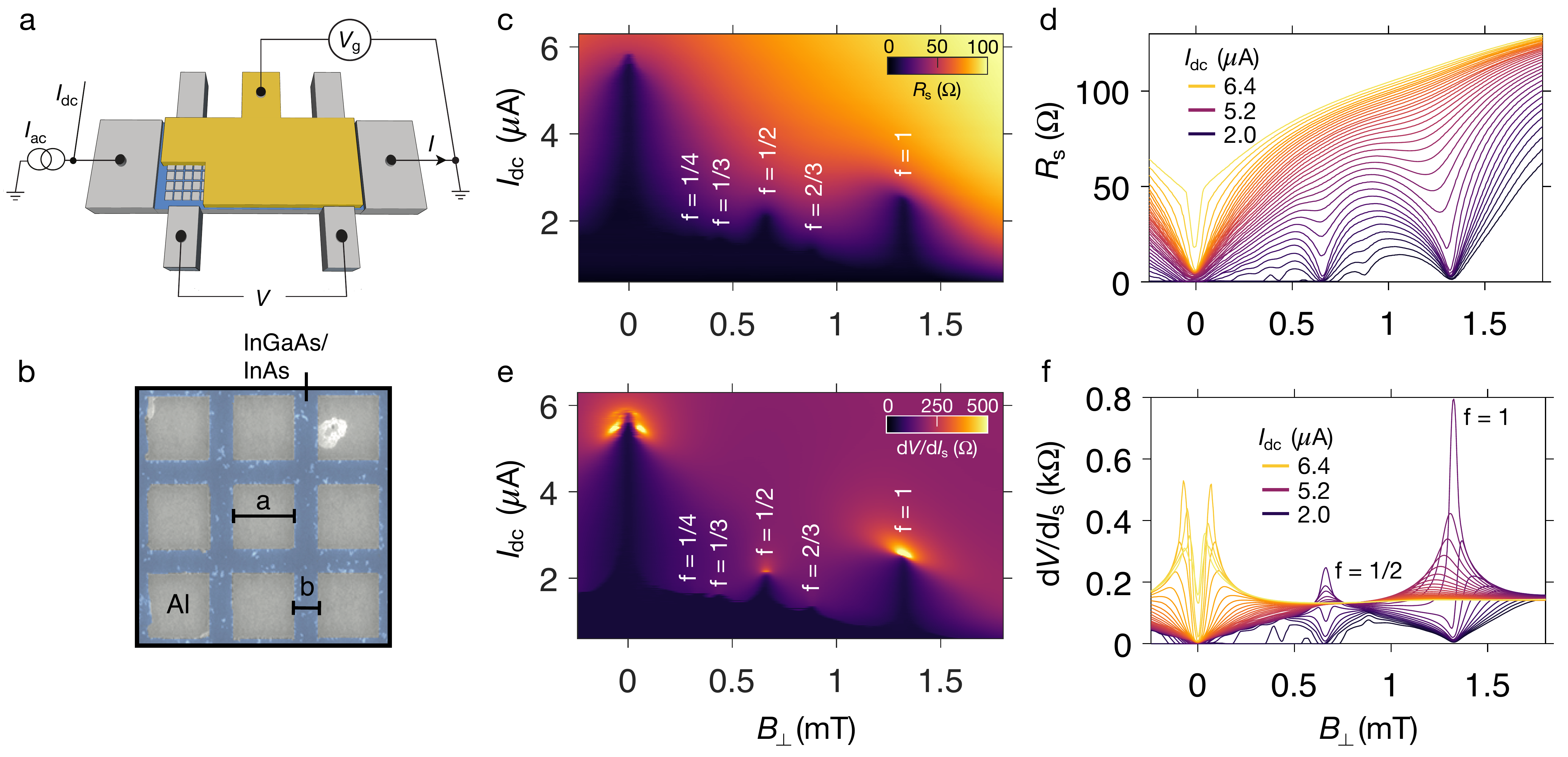}
	\caption{\textbf{Hybrid array and frustration-dependent dynamical vortex transitions}. \textbf{a,b)} Schematic of device with $W=100$ by $L=400$ epitaxial $a = 1\, \mu$m Al squares (gray), separated by $b=350$~nm strips (blue) of exposed InAs heterostructure, with ac + dc current bias, $I$, and voltage $V$ measured using side probes.  False-color micrograph taken before top gate was deposited. Gate voltage, $V_g$, controls carrier density in InAs strips between Al squares \cite{Boettcher2018}. \textbf{c)} Sheet resistance, $R_s \equiv (W/L) V/I$, as a function of dc current, $I_{\rm dc}$, and perpendicular magnetic field, $B_\perp$, shows $R_{s}=0$ for small $I_{\rm dc}$ with enhanced critical current, $I_{0}$ at $f=$~0, 1/4, 1/3, 1/2, 2/3, and 1. \textbf{d)} Line cuts of \textbf{c} shows dips in $R_s$ at  $f=$~0, 1/2, and 1, going to $R_{s} \sim 0$ for low $I_{\rm dc}$. \textbf{e-f)} Evolution of dips in differential resistance, $dV/dI_{s} \equiv (W/L)\ dV_{ac}/dI_{\rm ac}$, as a function of $B_{\perp}$  into a split peak at $f=0$, a symmetric peak at $f=1/2$, and an asymmetric peak at $f=1$. Each case is discussed in the text. }
	\label{fig1}
\end{figure*}

The device we investigated, shown in Fig.~\ref{fig1}, is based on an epitaxial Al/InAs heterostructure patterned by wet etching to form a square array of $1 \times 1\: \mu {\rm m}^{2}$ islands separated by 350~nm strips of exposed semiconductor \cite{Boettcher2018}. A Ti/Au top gate, separated from the array by 40 nm of atomic-layer deposited HfO$_2$ insulator, was used to control the carrier density in the strips between islands. The gated Hall bar has four side-probes for voltage measurements and two wide contacts at the ends for applying current, and is $W = 100$ islands wide with $L = 400$ between voltage probes. The sample is measured in a dilution refrigerator with a base temperature of 30 mK. All lines are filtered using QDevil rf and low-pass filters.

The top gate voltage, $V_g$, controls the Josephson coupling, $E_J$, between islands, driving the system from a superconducting state at $V_g \sim -3$~V to an insulating state at $V_g \sim -4$~V \cite{Boettcher2018}. The top gate also controls the barrier height, $E_B$, (energy saddle) between vortex pinning sites (energy minima) at the corners of the square islands. Numerical studies for metallic islands suggest $E_B \propto E_J$ \cite{Rzchowski1990}. Similar numerical studies have not been done for the semiconducting junctions. At intermediate gate voltages, an anomalous metallic phase was previously investigated, showing saturating gate-dependent sheet resistance at low temperature \cite{Boettcher2018, Boettcher.2022}. In this work, $V_g$ is set to yield a superconducting state at base temperature in the absence of applied dc current and was only modified slightly toward the anomalous metal phase where indicated.

 We probe the device in the nonlinear regime by measuring both the ac and dc parts of the total voltage, $V$ in a pair of side probe contacts, when total current $I$, consisting of a varied dc part, $I_{\rm dc}$ and ac part, $I_{\rm ac}=5$~nA, was applied to the end contacts. A magnetic field, $B_\perp$, was applied perpendicular to the plane of the array using an external solenoid controlled by a Keithley 2400 source-measurement unit. The area, $A=(a+b)^2$, of one plaquette of the array,  with $a=1\,\mu$m and $b=350$~nm, gives a characteristic magnetic field $\Phi_0/A = 1.4$~mT, where $\Phi_0 = h/2e$, corresponding to frustration $f=1$. Features associated with integer $f$ values are seen in Figs.~\ref{fig1}(c,d).

For small dc currents ($I_{\rm dc}\lesssim 2\, \mu$A), large dips in both sheet resistance, $R_s \equiv (W/L)\,V/I$, and differential sheet resistance, $dV/dI_s \equiv (W/L)\,dV/dI$, reach zero at $f=$~0, 1/2, and 1, with moderate dips at $f=$~1/4, 1/3 and 2/3, as seen in Fig.~\ref{fig1}(e). Increasing $I_{\rm dc}$ beyond an $f$-dependent critical value, $I_0(f)$, results in dip-to-peak transitions  in $dV/I_s$ while minima in $R_s$ remain minima, consistent with previous experiments~\cite{Poccia2015, Lankhorst2018, Mironov.2020, Rezvani.2020, Rezvani.2020b, Pei.2022}. The dip-to-peak transition in $dV/dI_s$, visible at $f=1/2$ and $f=1$, marks the onset of differential vortex motion without complete dissipative vortex flow. 

\section{Zero-field transitions from \\ superconductor and anomalous metal}
\label{split}

At zero magnetic field, $f=0$, the dynamical transition appears split, so that $dV/dI_s$ remains a minimum as a function of $B_\perp$, unlike $f=1/2$ and $f=1$, as seen in Figs.~\ref{fig1}(e,f).  The absence of a dip-to-peak transition at $f=0$, unlike for $f=$~1/2 and 1, is consistent with some previous results \cite{Poccia2015, Pei.2022} but not others \cite{Jiang.2004, Mironov.2020, Rezvani.2020b}, as discussed below.

We interpret the split peak as indicating that vortices are absent or sparse at $f=0$, consistent with a BKT picture in which unbound vortices pair and annihilate below a critical temperature, $T_{\rm BKT}$. 
This suggests that the zero-resistance state at $f=0$, where vortices are absent or sparse, is qualitatively different from the zero-resistance states at $f=1/2$ and $f=1$, where vortices of a single sign are abundant but frozen. 

The splitting of the transition around $f=0$ is found to increase at lower $I_{\rm dc}$, as seen in Figs.~\ref{fig1}(e,f). This is seen most clearly as the downward orientation of the bright features on either side of zero field in Fig.~\ref{fig1}(e), indicating that $I_0$ decreases with increasing vortex density, a signature of the role of vortex interaction in the dynamical transition.
This dependence is similar the vortex-density ($\propto f$) dependence of the vortex melting temperature \cite{Obaidat.2008}.

Tuning the gate voltage from $V_g=-2.00$~V  to $V_g=-3.23$~V drives the system from the superconducting state, where $R_{s}$ falls below measurement resolution, $\lesssim 0.1 \Omega$, at low temperature, to the anomalous metal phase, where $R_s$ saturates at a gate-voltage dependent value, up to $\sim h/4e^{2}$, at low temperature \cite{Boettcher2018, Boettcher.2022}. 

\begin{figure}[t]
	\includegraphics[width= 2.6 in]{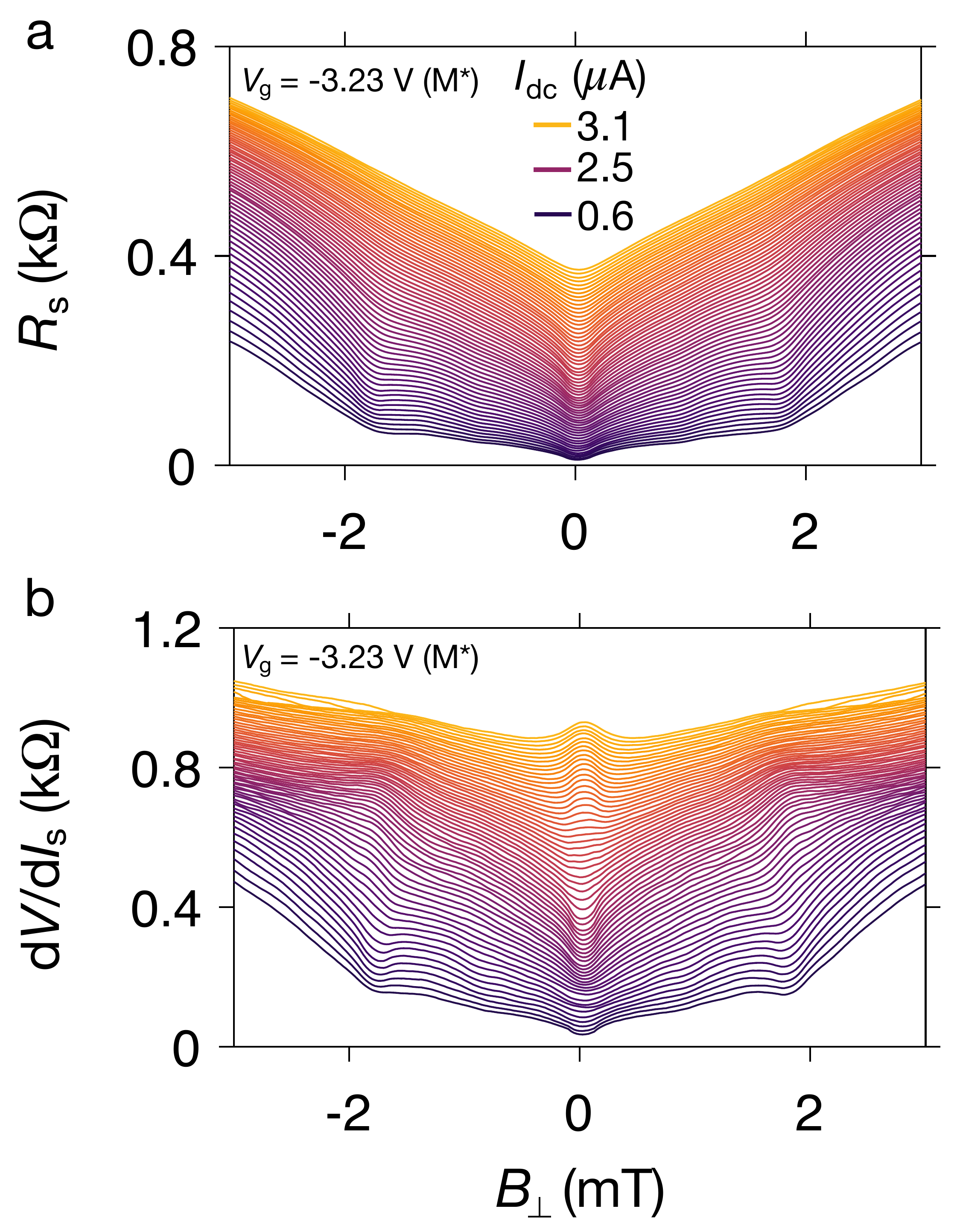}
		\caption{\textbf{Zero-field dynamical transition from the anomalous metal phase}. \textbf{a)} Sheet resistance, $R_{s}$, as a function of perpendicular magnetic field, $B_\perp$ over a range of dc currents, $I_{\rm dc}$. For all dc currents, zero field remains a resistance minimum, similar to the superconducting regime, Fig.~\ref{fig1}(d) \textbf{b)} Dip-to-peak transition at $f=0$ over the same range of dc currents. This behavior contrasts the superconducting regime in Fig.~\ref{fig1}(f), which shows a split peak at $B_\perp = 0$. The dip-to-peak transition suggests that vortices and antivortices are absent at $f=0$ in the superconducting phase but present in the anomalous metal phase.
			}
	\label{fig2}
\end{figure}

As shown in Fig.~\ref{fig2}, in the anomalous metal phase a dip-to peak-transition of $dV/dI_s$ as a function of $I_{\rm dc}$ {\it is} observed at $f=0$, while $R_s$ remains a minimum, similar to the dynamical transition observed at  $f=1/2$ and other integer $f$ values.

The {\it unsplit} transition in the anomalous metal phase suggests that vortices and antivortices (in equal number) are present at $f=0$, unlike in the superconducting phase. The nonvanishing $R_s$, which characterizes the anomalous metal phase, further suggests that at least some of the remnant vortices and antivortices are mobile even at $I_{\rm dc} < I_0$ and lowest temperatures. 

A picture of residual unpaired vortices at $f=0$ in the anomalous metal but not the superconductor is consistent with the observed vanishing of $T_{\rm BKT}$ in the anomalous metal \cite{Boettcher.2022}. One would anticipate that in the superconducting phase at higher temperatures, $T>T_{\rm BKT}$, an unsplit transition would occur at $f=0$. Reported dynamical transitions at $f=0$ are either at elevated temperatures \cite{Jiang.2004, Rezvani.2020b} or in the anomalous metal phase \cite{Mironov.2020}, consistent with our observations.

\section{Scaling and critical exponents}
\label{scaling}

The dip-to-peak transition at commensurate frustration can be described as melting of a frozen vortex lattice and has been analyzed as a dynamical Mott transition~\cite{Nelson1993, Poccia2015, Lankhorst2018, Granato.2018, Granato.2019, Mironov.2020, Rezvani.2020, Rezvani.2020b, Pei.2022}. Within this interpretation, one expects scaling at the transition among relevant parameters controlling the transition, namely the dc current bias, $I_{\rm dc}$, and the distance, $b \equiv f-f_{c}$, from the commensurate frustration values, in this case $f_{c}=$~1/2 or 1.

Following Refs.~\cite{Poccia2015, Lankhorst2018}, we introduce a scaling form for differential sheet resistance across the dip-to-peak transition, 
\begin{align}
dV(f, I)/dI_s - dV(f, I)/dI_s|_{I=I_{0}} = \mathcal{F}(|I - I_0|/|b|^{\varepsilon}),
\end{align}
\noindent where $\mathcal{F}(x)$ is a function of the scaled variable $ x \equiv |I- I_0|/|b|^{\varepsilon}$ and $\varepsilon$ is the scaling exponent. 

A scaling exponent $\varepsilon \sim 2/3$ was measured at $f=1$ \cite{Poccia2015, Lankhorst2018} and $f=2$ \cite{Poccia2015}, in a square array of Nb islands on Au, consistent with theory for a dynamical Mott transition. At $f=1/2$ in the same system, the value $\varepsilon \sim 1/2$ was found experimentally \cite{Poccia2015, Lankhorst2018}, and later confirmed numerically ~\cite{Granato.2018, Granato.2019}. Other experiments investigating the dynamic transition at commensurate $f$ found different exponents~\cite{Pei.2022} in the triangular lattice, or did not pursue scaling analysis~\cite{Mironov.2020, Rezvani.2020, Rezvani.2020b}. 

\begin{figure*}
	\includegraphics[width= 6
	in]{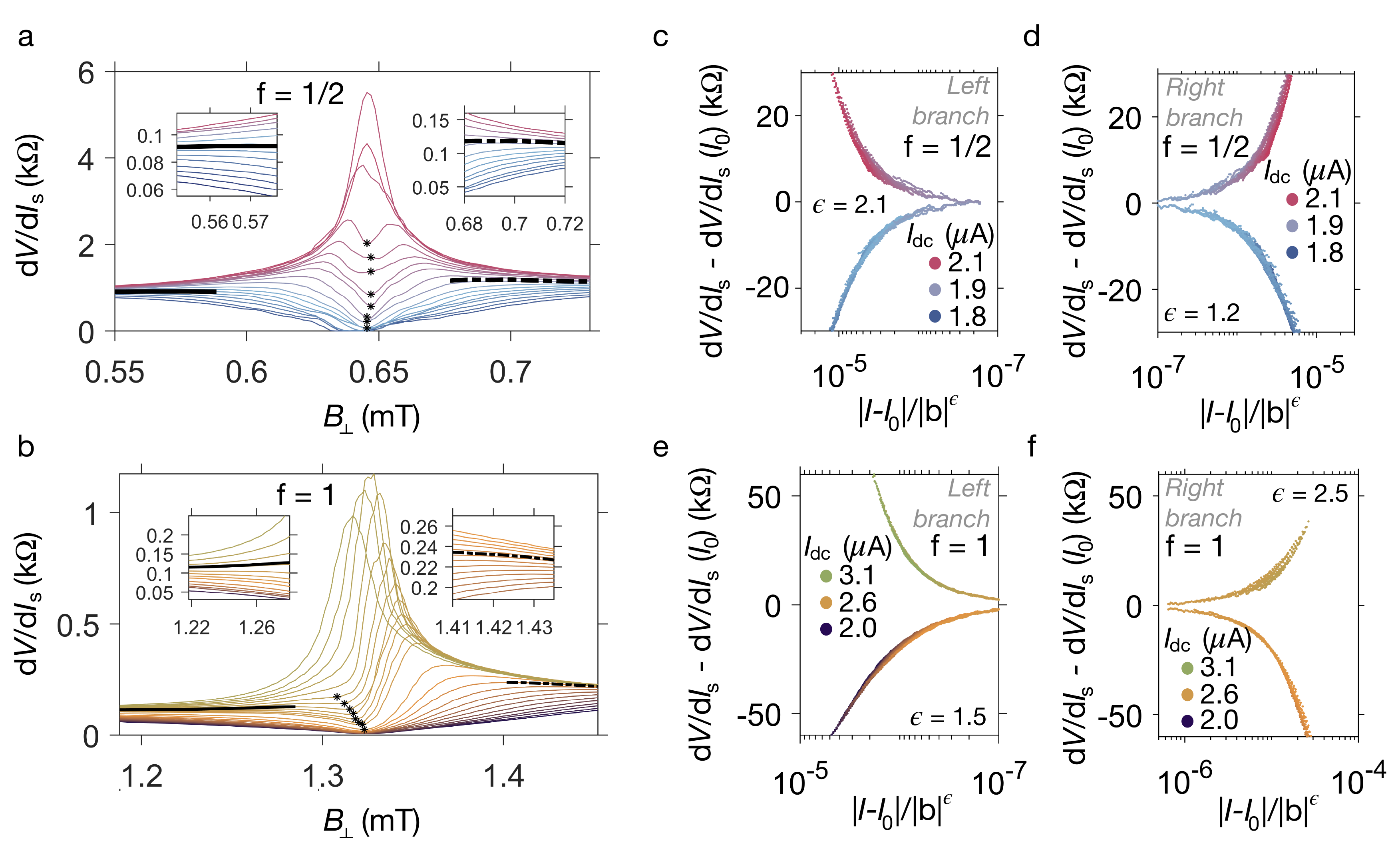}
		\caption{\textbf{Dynamical transitions, scaling, and critical exponents}. \textbf{a-b)} Differential sheet resistance, $dV/dI_s$, shows a dip-to-peak transition is a function of dc current, $I_{\rm dc}$, at commensurate frustration. ($f=1/2$) and full ($f=1$). Insets: the transitions on each side, denoted as the left and right branches of $f=1/2$ and $f=1$, are represented as superconductor-insulator transitions: down-bending curves are transitions to the pinned vortex state, while upward-bending curves are transitions to a state of vortex flow. The horizontal field-independent curves separating each state are marked as separatrices used in the scaling analysis (Left: black solid line. Right: black dotted line). Scaling plots of $f=1/2$ (\textbf{c-d}) and $f=1$ (\textbf{e-f}), showing that left and right branches yield different scaling exponents, i.e. there is an asymmetry around each critical frustration field. Exponents extracted for $f=1/2$ are $\epsilon =2.1$ (left) and 1.2 (right), while at $f=1$ we extract $\epsilon = 1.5$ (left) and 2.5 (right).
			}
	\label{fig3}
\end{figure*}

\begin{figure}
	\includegraphics[width= 2.6 in]{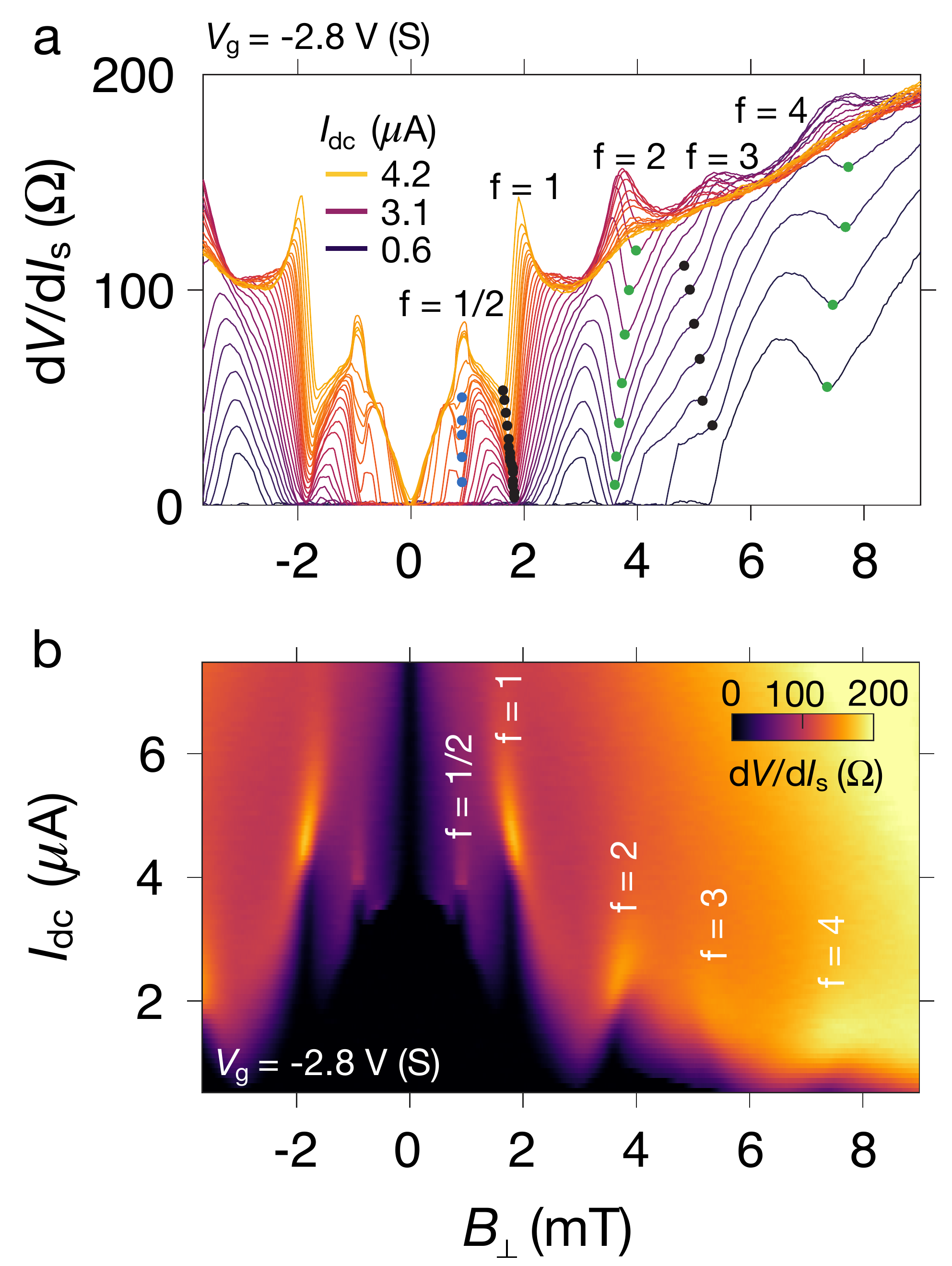}
		\caption{\textbf{Even, odd and zero flux states}. \textbf{a)} Differential sheet resistance curves for different values of dc current as a function of perpendicular magnetic field, $B_\perp$, at gate voltage $V_g=-2.8$~V, showing the evolution of vortex states for half-integer frustration $f=1/2$ and integers $f=1,2,3$ and 4. Odd minima (black dots) move down in field with increasing dc current, while even minima (green dots) move up in field. Minima at $f=1/2$ is roughly insenstive to dc current, consistent with the overall symmetry in $f$ of the $f=1/2$ transition. \textbf{b)} Differential sheet resistance (color) as a function of $B_\perp$ and $I_{\rm dc}$ shows commensurate features at factional and integer frustration, $f$. Bright features at the tips of the zero-resistance spikes are the dip-to-peak signature of the dynamical transition. Note the alternating, even-odd asymmetric bright features at integer $f$ values, the symmetric bright feature at $f=\pm 1/2$, and the vanishing differential sheet resistance remaining at large $I_{\rm dc}$ at $f=0$.
			}
	\label{fig4}
\end{figure}

The dip-to-peak transition at $f=1/2$ was found to be right-left symmetric, while the transition at $f=1$ was asymmetric, as discussed in Sec.~\ref{evenodd}. A consequence of the asymmetry is that the value of the commensurate frustration, $f_{c}$, depended on $I_{\rm dc}$ [dots in Figs.~\ref{fig3}(a,b)]. 
 The asymmetry at $f=1$ also yielded different separatrices on the left and right, $I_0^L = 2.5\: \mu\text{A}$, $I_0^R=2.35 \: \mu\text{A}$, while a single separatrix value was found for $f=1/2$ (solid and dashed curves in Figs.~\ref{fig3}(a,b). Similar asymmetries are also found at $f=$~2, 3 and 4, as seen in Fig.~\ref{fig4}.  
 
Scaling exponents $\varepsilon$ were obtained by fitting the slope on a log-log plot of differential resistance after subtracting the separatrix curve, $\frac{d}{dI}[dV/dI_{s}-dV/dI_{s}|_{I_0^{L/R}}]$ versus $1/b$ (see Methods). Scaled data collapse reasonably well, as seen in Figs.~\ref{fig3}(c-f), but yield different exponents on the left and right sides of the peaks, $\varepsilon = 2.1$ (left) and $\varepsilon =1.2$ (right) for $f=1/2$, and $\varepsilon = 1.5$ (left) and $\varepsilon =2.5$ (right)  for $f=1$, inconsistent with \cite{Poccia2015, Lankhorst2018}.

We speculate that different values for $\varepsilon$ could be caused by stronger pinning potential, $E_{B}$, in the Nb devices in \cite{Poccia2015,Lankhorst2018} compared to the Al array studied here, leading to a different regime where dynamics may be influenced by competing effects associated with binding potential and vortex-vortex interactions. A model of vortex pinning in metallic arrays yielded, $E_B = 0.2\,E_J$, where $E_J=(\hbar/2e)\,i_c$ is the Josephson coupling, and $i_c$ is the single-island critical current \cite{Rzchowski1990}. Taking $i_c\sim 5 \,\mu$A from \cite{Poccia2015} compared to $i_{c}\sim 0.5\,\mu$A in our Al array suggests roughly an order of magnitude difference in $E_B$.

\section{Skewed transitions and Even-odd structure}
\label{evenodd}
Figures \ref{fig1}(e) and \ref{fig2} reveal a striking difference between the dip-to-peak transitions at $f=1/2$ and $f=1$. Throughout the transition at $f=1/2$, each curve, representing a different $I_{\rm dc}$ value, is symmetric in $f$ about the transition point.  In contrast, for the $f=1$ transition, the curves are highly asymmetric, and in fact in the crossover from dip to peak are nearly antisymmetric lying above the high-current saturation on the right side and below it on the left side. This difference can also be seen in Fig.~\ref{fig1}(e). The bright feature at the top of the $f=1/2$ peak, corresponding to the peak in $dV(f)/dI_{s}$, is flat, while the bright feature at $f = 1$ is tilted, indicating that the peak in $dV(f)/dI_{s}$ occurs at lower $I_{\rm dc}$ at higher $f$. 

Symmetry in $f$ around $f=1/2$ follows from the presumed checkerboard vortex configuration near $f=1/2$ \cite{Teitel.1983}, which is symmetric with respect to the addition or subtraction of vortices. Excess or deficit vortices slightly above or below $f=1/2$ are expected to form a low-density superlattice on top of a base checkerboard \cite{Franz.1995, Lankhorst.2018b}. The symmetry between excess and deficit vortices should persist for weak disorder, which can then pin the superlattice. In contrast, $f=1$ is not symmetric with respect to the addition or subtraction of dilute vortices. Excess vortices above $f=1$ are repelled by each site in the full lattice, while deficit vortices, or antivortices, are attracted to each site, where they can annihilate with a vortex. Within this picture, excess vortices should be weakly pinned and contribute to low-current melting \cite{Franz.1995}, while vacancies resulting from annihilation with deficit vortices should more readily pin, lowering vortex mobility. The asymmetry between excess and deficit at $f=1$ is a signature of vortex interaction. 

Looking at integer transitions above $f=1$, an even-odd structure is evident in Fig.~\ref{fig4}, where the $f=3$ transition is skewed in the same direction as $f=1$, while transitions at $f=2$ and $f=4$ are skewed in the opposite direction. Even-odd behavior is also visible below the transition, in the current-dependent position of the minima of $dV/dI_{s}$, marked as dots in Fig.~\ref{fig4}(a). Some understanding of this structure follows from the arguments above, that half-filling of the array is a symmetry point, so one might expect $f=2$ to be show a reflection symmetry of the $f=1$ behavior about $f=3/2$, continuing by reflection about half-integers to higher integers.  

While the overall even-odd pattern at integer $f$ presumably reflects the square potential of the array, the difference between $f=1/2$ (symmetric) and $f=1$ (asymmetric) is also seen in triangular arrays \cite{Pei.2022}, and reflects a more basic difference between half and full filling. Even-odd structure of dynamical transition in triangular lattices has not been reported. It will be interesting to investigate vortex filling in various lattices experimentally and numerically.

\section{Discussion}
\label{discussion}

We have investigated a dynamical transition from frozen to mobile vortices in a gate-tunable superconductor-semiconductor Josephson junction array. Tuning the gate into the superconducting phase, where a zero-resistance state is observed at low temperature and current, we see dip-to-peak transitions in differential resistance near frustration $f=$~1/2, 1, 2, 3, and 4, similar to previous studies in metallic arrays. Motivated by the mapping of this transition to a Mott melting transition of frozen vortices, found good scaling at the transitions but not the Mott exponent found previously, perhaps due to a weaker binding potential in our Al arrays. 

The split transition at $f=0$ in the superconducting phase suggests that vortices are absent, not frozen, at $f=0$, consistent with a BKT model in which vortices and antivortices annihilate at $f=0$ below a critical temperature. When the array is tuned to the anomalous metal phase, a simple unsplit transition is observed at $f=0$ suggesting that frozen and perhaps some unfrozen vortices are present. These observations are consistent with previously reported experiments.

The transition at $f=1/2$ is symmetric in $f$ around the transition, reflecting the symmetry of the underlying half-filled checkerboard lattice. At $f = 1$, on the other hand, the transition is strongly asymmetric, suggesting that excess vortices on top of an underlying full lattice melt easily, but deficit vortices (antivortices) do not.  We find that the asymmetry of the $f=1$ transition persists to higher integers, mirrored about half integers, giving an overall even-odd pattern to transitions at higher integers. Further work is needed to understand this asymmetry, and how it depends on island and lattice geometry. This could also depend on the energetic of vortex configurations \cite{Berdiyorov.2005, Berdiyorov.2008} and binding multiple or giant vortices in the potential minima \cite{Baelus2002,Baelus2006}.

We emphasize an general connection between these highly tunable Josephson arrays and Bose-Hubbard systems \cite{Bruder.2005, Arovas.2021}. In these arrays, complexity is controlled by frustration and quantum by charging energy of the islands, relevant as the superconductor-insulator transition is approached. 

\emph{Acknowledgments} 
We thank A. Kapitulnik, S. Kivelson, B. Spivak, and V. Vinokur for useful discussions. 
Research supported by Microsoft Station Q, the Danish National Research Foundation, and a research grant (Project 43951) from VILLUM FONDEN.

\section*{Methods}
\label{methods}
\subsection*{Extracting scaling exponents}
This subsection provides details of the scaling analysis shown in Fig.~\ref{fig3}. The scaling exponent, $\varepsilon$, defined in Eq.~1 was  extracted from the slope of a log-log plot of $d/dI(dV/dI_s-dV/dI_s|_{I=I_0})$ versus $1/b$, allowing separate left and right critical currents, $I^L_0$ and $I^R_0$.  

Figure~\ref{figS} shows the spread of measured values. For each value of $b$ a mean of all data was calculated. A curve through the means is show as a solid black curve. Then, a line with offset was fit to the means. The linear fit is shown as a green dashed line. The slope of the linear fit yields $b$. The process is repeated for each value of $f$ and on the left and right sides for $f=1/2$ and $f=1$.
The value of $f$ extracted from the fit to $b$ takes into account the dependence of $f_c$ on $I_{\rm dc}$. Values of $f_c(I_{\rm dc})$ are shown as green and back dots in Fig.~\ref{fig4} 

We note that only points near the separatrix (above and below) were included in the analysis of extracting the scaling exponent for the right branch of $f_c=1$ due to the large asymmetry, see Fig.~\ref{figS}(d). 

\setcounter{figure}{0}
\makeatletter 
\renewcommand{\thefigure}{S\@arabic\c@figure}
\makeatother
\begin{figure}[t]
	\includegraphics[width= 3.4 in]{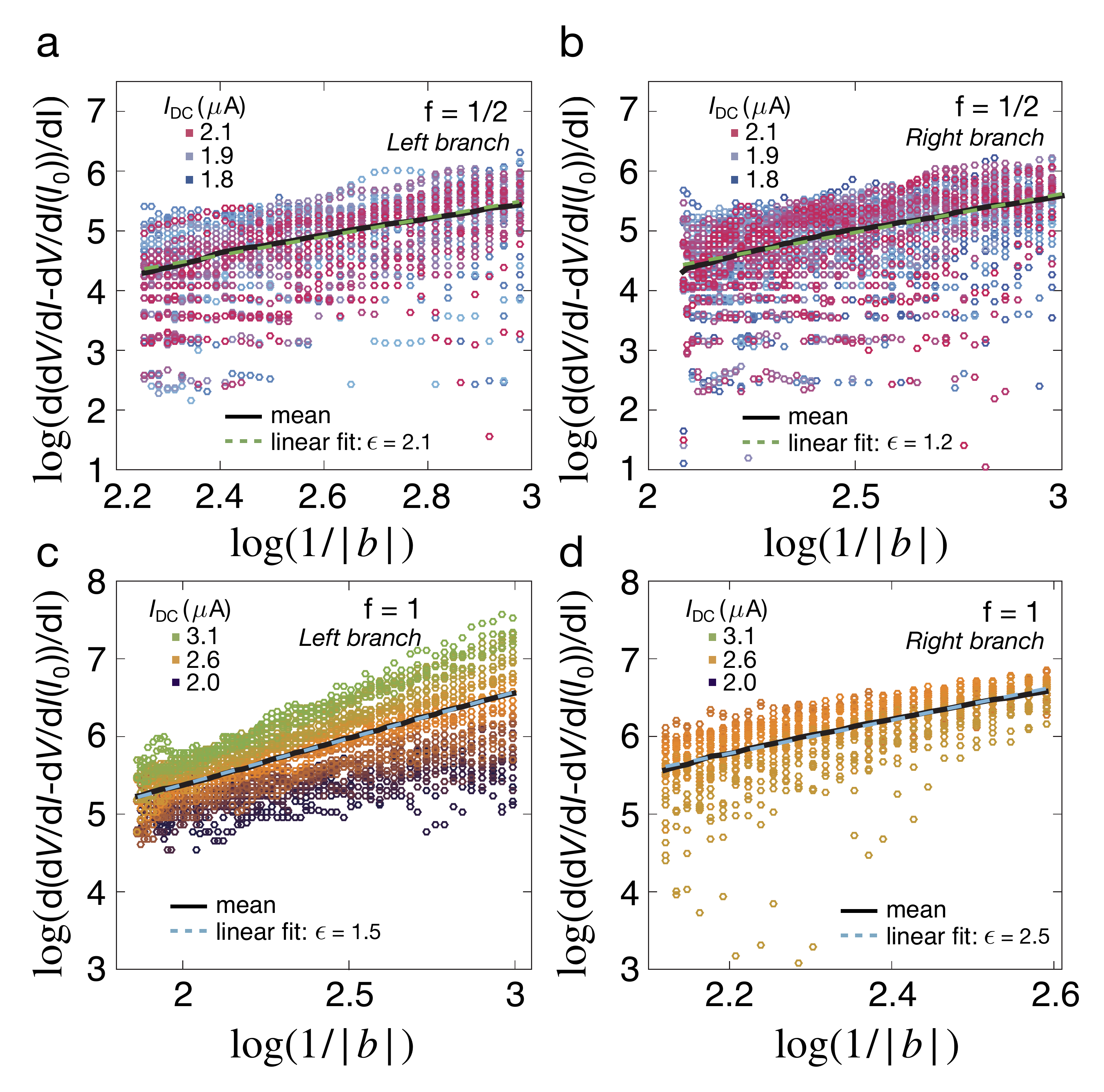}
		\caption{\textbf{Extracting scaling exponents}. \textbf{a-d)} Log-log plots of the slope of the differential resistance, with the separatrix subtracted, constructed to extract of scaling exponents, $\varepsilon$, separately for left and right branches around $f=1/2$ and $f=1$. Means of logs are calculated for each value of $b$ (solid black curve) then means fit to a linear function (dashed green line). }
	\label{figS}
\end{figure}

\newpage

\bibliography{Bibliography}

\end{document}